\documentclass[aps,prl,twocolumn,superscriptaddress,twoside,floatfix,amsmath,showpacs,nofootinbib]{revtex4}

\usepackage{graphicx}
\usepackage{amsmath}
\usepackage{textcomp}
\usepackage{dcolumn}% Align table columns on decimal point
\usepackage{bm}% bold math

\usepackage{amssymb}
\usepackage[latin1]{inputenc}
\usepackage{amsfonts}
\usepackage{lineno}
\usepackage{url}
\usepackage{array}
\usepackage{verbatim}
\usepackage{subfigure}
\usepackage{listings}
\usepackage{color}
\usepackage{lscape}
\usepackage{setspace}
\usepackage{footmisc}
\usepackage [english]{babel}
\usepackage [autostyle, english = american]{csquotes}
\MakeOuterQuote{"}

\DeclareMathSizes{10}{9}{6}{6}

\begin{document}

% \begin{frontmatter}

\title{New Concept for a Neutron Electric Dipole Moment Search using a Pulsed Beam}

\author{Florian~M.~Piegsa}
\email{florian.piegsa@phys.ethz.ch}
%\altaffiliation{Electronic addresses: florian.piegsa@phys.ethz.ch (FMP); zimmer@ill.fr (OZ).}
\affiliation{ETH Z\"urich, Institute for Particle Physics, CH-8093 Z\"urich, Switzerland}

\date{\today}

\begin{abstract}

A concept to search for a neutron electric dipole moment (nEDM) is presented, which employs a pulsed neutron beam instead of the nowadays established use of storable ultracold neutrons (UCN). The technique takes advantage of the high peak flux and the time structure of a next-generation pulsed spallation source like the planned European Spallation Source. It is demonstrated that the sensitivity for a nEDM can be improved by several orders of magnitude compared to the best beam experiments performed in the 1970's and can compete with the sensitivity of UCN experiments.

% A concept to search for a neutron electric dipole moment (nEDM) is presented, which takes advantage of the high peak flux and the time structure of next-generation pulsed spallation sources like the planned European Spallation Source. It is demonstrated that the sensitivity of a nEDM measurement can be improved by several orders of magnitude compared to the best beam experiments performed in the 1970's and can compete with the sensitivity of ultracold neutron experiments.

\end{abstract}

%\begin{keyword}
%Ultracold neutron production \sep Cryogenics \sep Fundamental neutron physics
%\end{keyword}

%\end{frontmatter}

%%%\twocolumn
%\linenumbers


% 07.20.Mc   Cryogenics Intrumentation
% 07.55.Ge 	 Magnetometers for magnetic field measurements 
% 11.30.Er   Charge conjugation, parity, time reversal, and other discrete symmetries 
% 14.20.Dh   Protons and neutrons 

% 95.36.+x   Dark Energy
% 03.65.-w   Quantum Mechanics
% 03.75.Be   Atom and neutron optics 

% 61.05.F-   Neutron diffraction and scattering
% 42.30.Rx   Phase retrieval
% 75.25.+z   Spin arrangements in magnetically ordered materials (including neutron and spin-polarized electron studies, synchrotron-source X-ray scattering, etc.)

% 34.20.Cf   Interatomic potentials and forces 
% 14.80.Va   Axions
% 14.70.Pw 	 Other gauge bosons 

% 13.40.Em 	 Electric and magnetic moments
% 07.55.Ge   Magnetometers for magnetic field measurements 
% 11.30.Er 	 Charge conjugation, parity, time reversal, and other discrete symmetries 
% 14.20.Dh 	 Protons and neutrons  (Particle Properties)

% 67.85.-d   Ultracold gases

% 29.25.Dz 	 Neutron sources 
% 28.20.Fc 	 Neutron absorption 
% 28.20.Cz   Neutron Scattering
% 78.70.Nx 	 Neutron inelastic scattering 

\maketitle

% \section{Introduction}
%
%
%
\begin{figure*}
	\centering
		\includegraphics[width=0.90\textwidth]{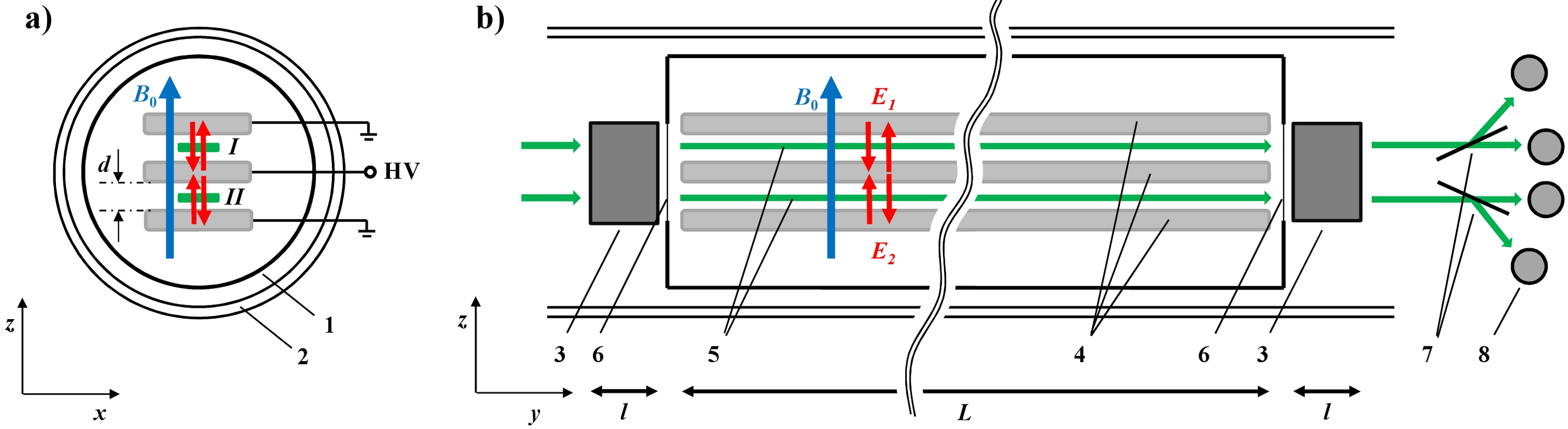}
	\caption{(color online). Schematic drawings of the proposed nEDM pulsed beam experiment. (a) Cross section of the experiment with two separated neutron beams directed along the $y$-axis (\textit{I} and \textit{II}) located between three electrodes and in a static magnetic field $B_0$: (1) vacuum flight tube and (2) several layers of mu-metal for passive magnetic shielding. An actively stabilized system of surrounding compensation coils might be advantageous (not shown). 
(b) Longitudinal cut of the experimental setup: (3) two $\pi/2$ spin-flip coils of length $l$, (4) high voltage electrodes of length $L$ providing vertical electric fields $E_1$ and $E_2$, (5) two neutron beams, (6) aluminum vacuum beam windows, (7) polarization analyzing supermirrors and (8) neutron detectors.}
	\label{fig:Setup}
\end{figure*}
The search for electric dipole moments of fundamental particles and atoms presents a very promising route for finding new physics beyond the Standard Model of particle physics \cite{Raial/2008, Khriplovich/1997}. A permanent electric dipole moment violates parity (P) and time reversal symmetries (T) and, invoking the CPT theorem, also CP symmetry. However, new sources of CP violation are expected to be found in order to understand the observed large matter-antimatter asymmetry in the universe \cite{Sakharov/1967,Riotto/1999,Pospelov/2005} and because most extensions of the Standard Model allow for new CP violating phases. \\
Already in 1950, Purcell and Ramsey proposed a scheme to search for a non-vanishing neutron electric dipole moment $d_{\text{n}}$ (nEDM) \cite{Purcell/1950}. 
An upper limit for $d_{\text{n}}$ is derived by comparing the neutron Larmor precession frequencies in a constant magnetic field $B_0$ superimposed with an electric field $E$ applied parallel and anti-parallel to $B_0$, respectively. The difference in precession frequency is given by
\begin{equation}
  \hbar \Delta\omega = (-2\mu_{\text{n}} B_0 - 2 d_{\text{n}} E)-(-2\mu_{\text{n}} B_0 + 2 d_{\text{n}} E) = -4 d_{\text{n}} E
\end{equation}
where $\hbar$ is Planck's constant and $\mu_{\text{n}}$ is the magnetic moment of the neutron. In this formula the critical assumption is made that the magnetic field $B_0$ does not change during the course of the two measurements. Historically, the early nEDM experiments have been performed using neutron beams \cite{Smith/1957,Miller/1967,Baird/1969,Dress/1977,Ramsey/1986}, while current experiments and new projects prefer using ultracold neutrons (UCN) \cite{Baker/2006,Altarev/1996,Grinten/2009,Masuda/2012a,Ito/2007,Altarev/2012,Baker/2011,Serebrov/2009b,Lamoreaux/2009}.\footnote{A complementary approach to measure the nEDM using a neutron beam and crystal diffraction has so far reached a sensitivity of $6.5\times10^{-24}$~e$\cdot$cm \cite{Federov/2009}.} Both methods employ Ramsey's Nobel prize winning molecular beam method of separated oscillatory fields adapted to neutrons \cite{Ramsey/1949,Ramsey/1950} to measure the neutron spin precession phase $\varphi=\Delta\omega \cdot T$. Here, $T$ is the interaction time of the neutron spin with the applied electric field. Experiments with UCN have the eminent advantage of much longer interaction times (in the order of $100$~s compared to about $10$~ms for neutron beam experiments), since UCN can be confined in so-called neutron bottles made of suitable materials with small loss cross sections \cite{Golub/1991}. 
This results in a largely improved sensitivity, since the statistical uncertainty (standard deviation) on the nEDM can be derived as 
\begin{equation}
\label{deltadn}
  \sigma(d_{\text{n}}) = \frac{\hbar}{2 \eta  T E \sqrt{N}}  
\end{equation}
where $N$ is the total number of detected neutrons and $0 \leq \eta \leq 1$ the "visibility" of the Ramsey fringe pattern \cite{Golub/1972}. By contrast, much larger neutron count rates and up to $10$ times higher electric fields can be achieved in neutron beam experiments \cite{Baumann/1988,Dress/1977,Baker/2006}. The latter is possible because neutron beams do not require insulating wall material mounted between the high voltage electrodes as in experiments with stored UCN.    % (leakage currents). \\
However, the limiting systematic effect in beam experiments has so far been the relativistic $v \times E$-effect, which arises from the motion of the neutron through the electric field producing an effective magnetic field $\vec{B}_{v \times E} = -(\vec{v} \times \vec{E}) / c^2$ according to Maxwell's equations, with $c$ being the speed of light in vacuum. In the most recent nEDM beam experiment this effect was corrected for by mounting the entire Ramsey spectrometer on a turntable, in order to reverse the direction of the neutron beam with respect to the apparatus \cite{Dress/1977}.
% \footnote{A similar technique is applied in experiments searching for an electron electric dipole moment using atomic beams of $^{205}$Tl, where the $v \times E$-effect is compensated by measuring with counterpropagating beams traveling vertically up and down \cite{Commins/1994,Regan/2002}.}
Stored UCN, however, have an average velocity of approximately zero and therefore the $v \times E$-effect is substantially reduced, seemingly rendering nEDM experiments with beams obsolete. \\
Lately, several sensitive Ramsey experiments using neutron beams have been performed \cite{vdBrandt/2009,Piegsa/2009a,Piegsa/2008b,Piegsa/2009b,Piegsa/2011,Piegsa/2012a,Piegsa/2012b}, which revived the previously abandoned idea of a nEDM beam experiment. 
Here, a concept is presented which overcomes the drawback and is able to reach sensitivities of UCN experiments. This is achieved by directly measuring the $v \times E$-effect  
by employing a high intensity pulsed neutron beam. Such beams will be made available in the near future at % next-generation spallation sources like 
the planned European Spallation Source (ESS) \cite{ESS/web} or possibly at Fermilab's Project X \cite{Projectx/web}.  \\

%
%
% \section{Experimental Apparatus}
%
In Fig.\ \ref{fig:Setup} a scheme of the experimental setup of the proposed nEDM beam concept is presented. 
Two separated neutron beams ($I$ and $II$) with a cross section of several cm$^2$ and with the velocity directed along the $y$-axis are initially polarized in $z$ direction. They are traveling inside a non-magnetic vacuum flight tube to avoid neutron scattering and absorption in air. Several layers of mu-metal provide shielding from the Earth's magnetic field and other disturbing magnetic field sources. The beams are exposed to a static and homogeneous magnetic field $B_0$ and electric fields $E_1$ and $E_2$ applied along the $z$-axis. In principle, the magnitude of $B_0$ can be chosen arbitrarily. For practical reasons and the suppression of systematic effects, however, a field of about $200$~$\mu$T seems reasonable. % \footnote{One reason is the length $l$ of the neutron $\pi/2$ spin-flip coils, for which $l \gtrsim 2\pi v/\omega_0$ should hold.}
The electric fields are established by means of three horizontally oriented parallel metallic electrodes (e.g.\ made from aluminum) with a total length $L \approx 50$~m and a distance $d$ of some centimeters. 
The electrodes might be assembled from many well aligned short sections of $1$~m length. 
A horizontal electrode geometry is preferable, since neutrons of all velocities experience the same magnetic field, in contrast to a vertical arrangement where slow and fast neutrons will describe different flight parabola due to the gravitational interaction. 
Depending on the polarity of the high voltage applied to the middle electrode (the outer electrodes are connected to ground) the electric fields are oriented anti-parallel/parallel or parallel/anti-parallel with respect to $B_0$. In order to avoid large losses due to beam divergence in $z$ direction the electrodes can be coated with a non-depolarizing supermirror multilayer structure, e.g.\ Cu/Ti or NiMo/Ti \cite{Padiyath/2004,Schanzer/2009}. Instead of metallic electrodes, one could alternatively employ neutron guide float glass utilizing the metallic supermirror coating as a thin conducting electrode layer. 
The Ramsey setup consists of two $\pi/2$ spin-flip coils which produce phase-locked oscillatory fields perpendicular to $B_0$, e.g.\ longitudinal in $y$ direction. They are driven with a frequency $\omega_{\text{RF}}$ close to the neutron Larmor precession frequency $\omega_0 = - \gamma_{\text{n}} B_0$, where $\gamma_{\text{n}}$ is the gyromagnetic ratio of the neutron. The amplitudes of the oscillatory fields need to be modulated in time and synchronized with the repetition rate of the spallation source, in order to produce optimal $\pi/2$ flips for neutrons of all velocities $v$ present in a neutron pulse \cite{Maruyama/2003}. Between the spin-flip coils, the neutron spins precess in the $x$-$y$ plane perpendicular to the externally applied fields. The spins of the neutrons are analyzed by polarizing supermirrors which are transparent for one spin state and reflect the other and, thus, allow to separately detect both spins species. 
\begin{figure}
\centering
	\includegraphics[width=0.45\textwidth]{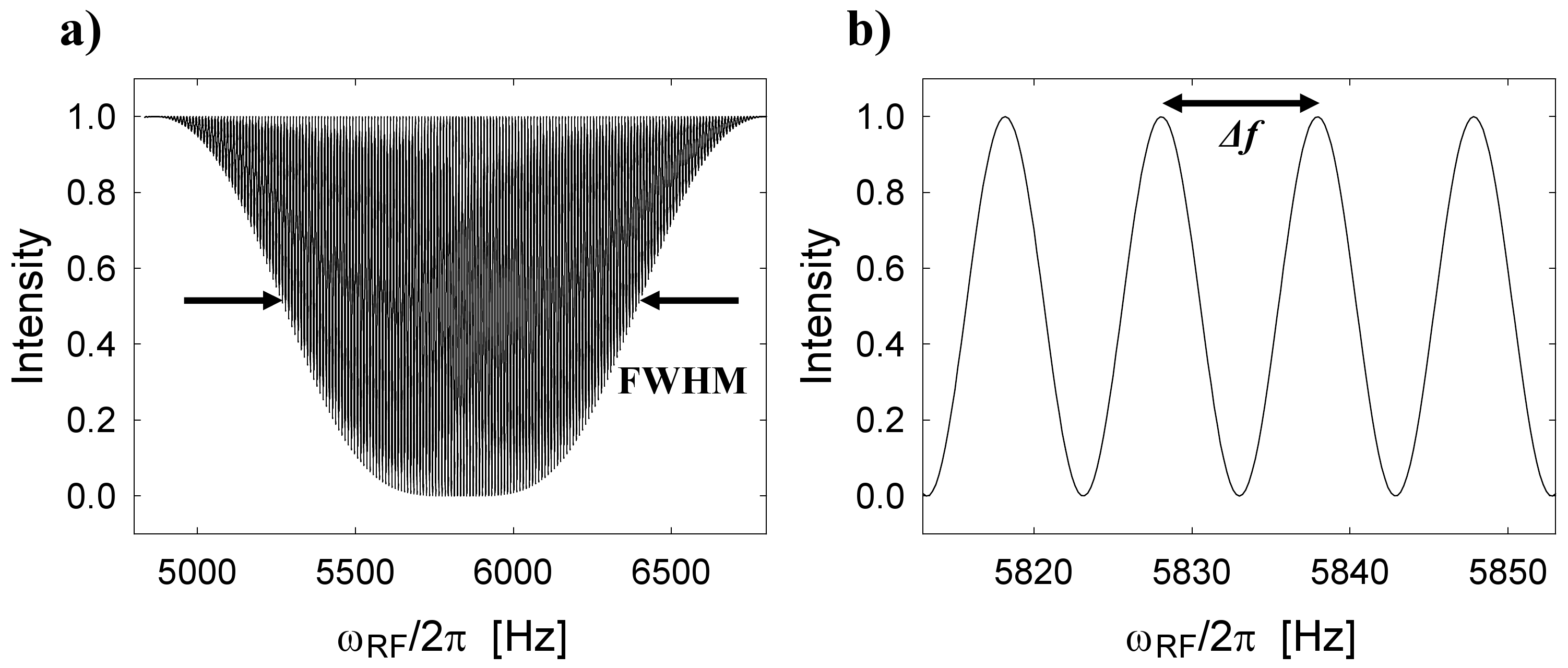}
  \caption{Simulated Ramsey pattern as a function of $\omega_{\text{RF}}$ for neutrons with a velocity $v=500$~m/s and $L=50$~m, $l=0.5$~m, $\eta=1$ and $B_0 = 200$~$\mu$T, i.e.\ $\omega_0/2\pi \approx 5833$~Hz. (a) Complete Ramsey pattern with FWHM~$\approx 1.1$~kHz. (b) Center fringes of the pattern with $\Delta f \approx 10$~Hz.}
\label{fig:RamseySim}
\end{figure} 
The neutrons are detected as a function of time-of-flight in four detectors capable standing high count rates, compare e.g.\ \cite{Klein/2011,Gledenov/1994}. A so-called Ramsey pattern is obtained by measuring the count rate as a function of $\omega_{\text{RF}}$. One obtains Ramsey patterns for each beam and each time-of-flight bin, i.e.\ each neutron velocity. In Fig.\ \ref{fig:RamseySim} a simulated signal is presented. The distance between two neighboring fringe maxima is given by $\Delta f \approx 1/T = v/L$ and the width of the envelope of the pattern is $\text{FWHM} \approx 1.12 \cdot v/l$, where $l$ is the length of the spin-flip coils \cite{Piegsa/2008a}.\footnote{Instead of performing a scan of the frequency, a similar Ramsey pattern is obtained by scanning the relative phase between the two oscillatory fields, which allows that the resonance condition $\omega_{\text{RF}}=\omega_0$ is fulfilled permanently.} Any additional precession of the neutron spins between the two $\pi/2$ spin-flip coils, for instance due to a nEDM, will cause a corresponding phase shift of the Ramsey fringes. The use of two beams allows to correct for phase drifts of the Ramsey patterns which equally appear in both beams (common noise rejection).\\
An aspect which needs to be taken into account, is dephasing of the neutron spins during precession. In order to avoid the accompanied loss in visibility of the Ramsey fringes, the lateral magnetic field gradients of $B_0$ averaged over the flight path need to be limited. The gradients should not exceed $2 \cdot (\partial B_0/ \partial x) \approx (\partial B_0 / \partial z) \lesssim \pi / (4 \gamma_{\text{n}} T d) \approx 15$~nT/cm, with $d=3$~cm and $T=0.1$~s. Such a field uniformity is achieved in the center of a Helmholtz coil with a radius of about $40$~cm. Hence, the $B_0$ field can be provided with a long pair of rectangular coils or a $\cos{\theta}$-coil with comparable dimensions. 
Furthermore, magnetic field and field gradients should be constantly monitored with an array of sensors, e.g.\ fluxgates and atomic magnetometers \cite{Groeger/2006,Knowles/2009,Green/1998,Gemmel/2010}. An intriguing possibility would be to integrate a co-propagating beam of polarized $^3$He atoms as a magnetometer/gradiometer ($\gamma_{\text{$^3$He}}/ \gamma_{\text{n}} \approx 1.1$) \cite{Eckel/2012,DeKieviet/1995}. Alternatively, two additional neutron beams, which are not exposed to the electric fields, traveling below and above the two nEDM beams would serve the same purpose.  
\begin{figure}
	\centering
		\includegraphics[width=0.450\textwidth]{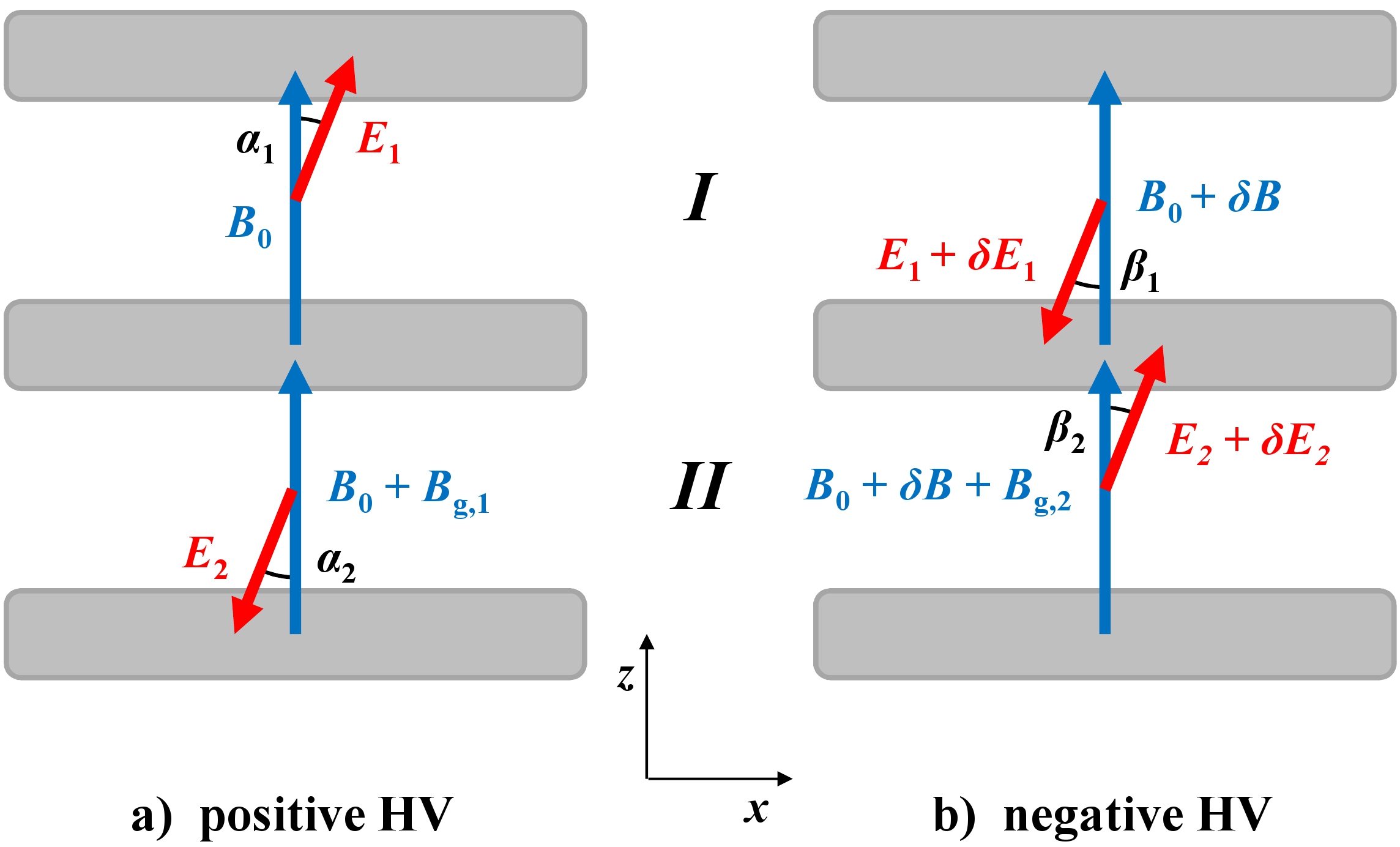}
	\caption{(color online). Electric and magnetic fields experienced by the two neutron beams ($I$ and $II$) for the two cases: (a) positive and (b) negative high voltage applied to the middle electrode.}
	\label{fig:Electrodes}
\end{figure}   \\

%
%
% \section{Measurement Principle}
%
The described Ramsey setup represents a very sensitive apparatus to measure small magnetic and pseudomagnetic fields very accurately by determining the phase shifts of the Ramsey fringes. A nEDM interacting with an electric field can be described by the pseudomagnetic field $B^*=2 d_{\text{n}} E / (\hbar \gamma_{\text{n}})$. Inserting the present best upper limit $d_{\text{n}}=2.9 \times 10^{-26}$~e$\cdot$cm (90\% C.L.) \cite{Baker/2006} and an electric field of $10$~MV/m yields a corresponding field $B^* \approx 50$~fT. The magnitude of a magnetic field due to the $v \times E$-effect for neutrons with a velocity of $500$~m/s in the same electric field is many orders of magnitude larger $B_{v \times E} \approx 55$~nT. Firstly, however, in a nEDM experiment $B_{v \times E}$ is oriented perpendicular to the main field $B_0$ and thus leads only to a small correction 
and secondly, it is proportional to the neutron velocity. Hence, by employing a pulsed neutron beam and measuring Ramsey patterns for different velocities the effect on the neutron spins caused by $B^*$ and $B_{v \times E}$ can be well separated.  \\
% In order to understand the gist of the new concept we consider Fig.\ \ref{fig:Electrodes}, which depicts an overview of the electric and magnetic fields experienced by the neutron beams $I$ and $II$.
% In Fig.\ \ref{fig:Electrodes} an overview is given depicting the electric and magnetic fields experienced by the neutron beams $I$ and $II$.
The electric and magnetic fields experienced by the neutron beams $I$ and $II$ are depicted in Fig.\ \ref{fig:Electrodes}.
A nEDM is determined by two Ramsey measurements with different electric field settings, here achieved by applying either a positive or negative high voltage to the middle electrode. In this generalized scheme also a non-perfect alignment of the fields, field instabilities and magnetic field gradients are taken into account. The effective magnetic fields for a positive voltage are given by
\begin{eqnarray}
   \vec{B}_{I,+} & = & \left( \begin{array}{c} -\frac{v E_1}{c^2} \cos{\alpha_1} \\  0  \\ \frac{v E_1}{c^2} \sin{\alpha_1} + B_0 + B^*  \end{array} \right)                \label{positiveHV1} \\
   \vec{B}_{II,+} & = & \left( \begin{array}{c} \frac{v E_2}{c^2} \cos{\alpha_2}  \\  0  \\ -\frac{v E_2}{c^2} \sin{\alpha_2} + B_0 + B_{g,1} - B^*  \end{array} \right)    \label{positiveHV2}
\end{eqnarray}
And for a negative applied voltage
\begin{eqnarray}
  \vec{B}_{I,-}  & = & \left( \begin{array}{c} \frac{v (E_1 + \delta E_1)}{c^2} \cos{\beta_1} \\  0  \\ -\frac{v (E_1 + \delta E_1)}{c^2} \sin{\beta_1} + B_0 + \delta B - B^* \end{array} \right) \label{negativeHV1} \\
  \vec{B}_{II,-} & = & \left( \begin{array}{c} -\frac{v (E_2 + \delta E_2)}{c^2} \cos{\beta_2} \\  0  \\ \frac{v (E_2 + \delta E_2)}{c^2} \sin{\beta_2} + B_0 + \delta B + B_{g,2} + B^* \end{array}  \right) \label{negativeHV2}
\end{eqnarray}
where the magnetic fields $B_{g,1}$ and $B_{g,2}$ represent magnetic field gradients in $z$ direction. In Eq.\ (\ref{positiveHV1}) - (\ref{negativeHV2}), we have assumed small tilting angles $\alpha_{i}$ and $\beta_{i}$ for $i \in \{1,2\}$, negligible changes of the electric field magnitudes after polarity reversals, i.e.\ $\delta E_{i} \ll E_{i}$, and $E_1 \approx E_2$, to approximate the pseudomagnetic fields due to the nEDM in all cases by $B^* =  2 d_{\text{n}} E_1 / (\hbar \gamma_{\text{n}})$. 
Magnetic fields in $y$ direction are neglected since the only case where they become relevant, namely a geometric phase, is treated later.
Further, by taking into account that the magnetic field change $\delta B$ between the two measurements, $B_{g,1}$, $B_{g,2}$ and $B^*$ are all much smaller than $B_0$, one derives the frequency shift
%\begin{widetext}
%\begin{eqnarray}
%\label{magnfieldmagnitudes}
%  |\vec{B}_{I,+}|   & \approx & (B_0 + B^*) + \left(\frac{v E_1}{c^2}\right) \sin{\alpha_1} + \frac{1}{2 B_0} \left(\frac{v E_1}{c^2}\right)^2 \\
%  |\vec{B}_{II,+}|  & \approx & (B_0 + B_{g,1} - B^*) + \left(\frac{v E_2}{c^2}\right) \sin{\alpha_2} + \frac{1}{2 B_0} \left(\frac{v E_2}{c^2}\right)^2 \\
%  |\vec{B}_{I,-}|   & \approx & (B_0 + \delta B - B^*) + \left(\frac{v (E_1 + \delta E_1)}{c^2}\right) \sin{\beta_1} + \frac{1}{2 B_0} \left(\frac{v (E_1+\delta E_1)}{c^2}\right)^2 \\
%  |\vec{B}_{II,-}|  & \approx & (B_0 + \delta B  + B_{g,2} + B^*) + \left(\frac{v (E_2 + \delta E_2)}{c^2}\right) \sin{\beta_2} + \frac{1}{2 B_0} \left(\frac{v (E_2+\delta E_2)}{c^2}\right)^2
%\end{eqnarray}
%\end{widetext}
% Hence, we finally arrive at
\begin{eqnarray}
  \Delta \omega & = &    \gamma_{\text{n}} \cdot  \left(  | \vec{B}_{I,+} | - | \vec{B}_{II,+} |  -  | \vec{B}_{I,-} |   +  | \vec{B}_{II,-}|  \right)   \label{eq:magnfielftot1} \\
                & \approx & \gamma_{\text{n}} \cdot \left(  4 B^* + \delta B_g + \left(\frac{v E'}{c^2}\right)  + \frac{1}{B_0} \left(\frac{v E''}{c^2}\right)^2    \right) \label{eq:magnfielftot2}
\end{eqnarray}
using that $(\delta E_i)^2 \ll E_i \delta E_i$ and with $\delta B_g = ( B_{g,2} - B_{g,1} )$ describing the change in the magnetic field gradient. $E' = E_1 \sin{\alpha_1} + E_2 \sin{\alpha_2} + (E_1 + \delta E_1) \sin{\beta_1} + (E_2 + \delta E_2) \sin{\beta_2} $ and $(E'')^2 = ( E_2 \delta E_2 - E_1 \delta E_1 ) + ( E_2^2 \alpha_2 \delta\alpha_2 - E_1^2 \alpha_1 \delta\alpha_1 )$, with $\delta \alpha_i=\alpha_i - \beta_i$. Hence, the velocity dependent phase shift 
\begin{equation}
\label{phases}
\varphi (v) =  \gamma_{\text{n}} L \cdot \left( \left( \frac{4 B^* + \delta B_g}{v} \right) + \frac{E'}{c^2} + \frac{v}{B_0}\left(\frac{E''}{c^2}\right)^2 \right)
\end{equation}
can be divided into three parts. The first part containing the pseudomagnetic nEDM effect and $\delta B_g$ is proportional to $1/v$, while the first and second order terms of the $v \times E$-effect are constant and proportional to $v$, respectively. If the second order term is sufficiently suppressed, a value or upper limit for the nEDM can be extracted by plotting $\varphi(v)$ as a function of $1/v$ and determining the slope by a linear fit.\\

%
%
%
% \section{Statistical Sensitivity}
%
In the following, the statistical sensitivity of the proposed concept is compared to experiments using UCN. Since the uncertainty on $d_{\text{n}}$ given in Eq.\ (\ref{deltadn}) scales with $1/T$, the length of the spin precession region should be as large as possible. In the neutron-antineutron oscillation experiment performed at the research reactor of the Institut Laue-Langevin (ILL) in Grenoble, an approximately $75$~m long neutron flight tube shielded with mu-metal was employed \cite{Baldo/1994,Bitter/1991}. Assuming a similar setup with $L=50$~m and neutrons with an average velocity of 500~m/s yields $T=0.1$~s. Thus, a gain factor $f_{T}=0.001$ can be expected, since precession times of about 100~s can be routinely achieved in UCN experiments. Due to the higher electric fields reachable in beam experiments, about $5-10$~MV/m instead of $1$~MV/m, a gain factor $f_{E}=5-10$ is obtained. 
The neutron count rate in the latest UCN experiment was approximately $60$~s$^{-1}$, as an average of $14000$~UCN were detected per $240$~s long measurement cycle \cite{Baker/2006}.
At the planned spallation source ESS the time averaged flux will be equivalent to the continuous flux at the ILL reactor with an unpolarized neutron capture flux-density comparable to the one available at the fundamental physics beam line PF1b of about $2 \times 10^{10}$~cm$^{-2}$s$^{-1}$ \cite{Peggs/2012, Abele/2006}.
The ESS will produce pulses of approximately $3$~ms length with a repetition rate of about $14$~Hz. Hence, assuming a total source-to-detector distance of $75$~m, allows for neutron velocities e.g.\ between $660$~m/s and $400$~m/s, i.e.\ a neutron de Broglie wavelength band from $0.6$~nm to $1.0$~nm. This band can be selected most efficiently with only small losses, by using neutron optical devices installed upstream of the experimental setup. For instance by means of a frame overlap filter to scatter out neutrons with longer wavelengths and a curved neutron guide to avoid transmission of neutrons with a wavelength shorter than $0.6$~nm.
Integrating the differential flux-density given in Ref.\ \cite{Abele/2006} over this wavelength range yields a neutron particle flux-density of $1.5 \times 10^{9}$~cm$^{-2}$s$^{-1}$. Thus, employing a polarizing cavity with an average transmission of $30-35$\%, a polarized neutron particle flux-density of about $5 \times 10^{8}$~cm$^{-2}$s$^{-1}$ is deduced \cite{Aswal/2008,SwissN/web}.
Together with an estimated correction of $4 \times 10^{-4}$ or $2\times 10^{-2}$ for divergence losses in two dimensions (with absorbing electrodes) or one dimension (with supermirror coated electrodes), respectively, one can expect a neutron count rate at the detector between $2 \times 10^5$~cm$^{-2}$s$^{-1}$ and $1 \times 10^7$~cm$^{-2}$s$^{-1}$. 
These values are consistent with the measured unpolarized neutron flux of about $10^7$~cm$^{-2}$s$^{-1}$ after almost 100~m free propagation given in Ref.\ \cite{Baldo/1994} and calculations using flux brightness data from Ref.\ \cite{Abele/2006}. 
This leads to a gain factor of $f_{\sqrt{N}} \approx 360-2600$ for two neutron beams with a cross section of 20~cm$^2$ each (e.g.\ $7\times3$~cm$^2$). 
Ultimately, one can further improve the sensitivity by at least a factor $f_{u} \approx \sqrt{2}$, by means of a neutron optical system with parabolic/elliptic guides focusing on the cold moderator \cite{Schanzer/2004}. %\footnote{A potential additional gain factor consists in employing a polarizing beam switch feeding the two beams with neutrons of opposite spin states, as proposed in \cite{Cho/2009}.}
In total, this results in a sensitivity gain $f_{T} f_{E} f_{\sqrt{N}} f_{u} \approx 2.5 - 40$ compared to the present best UCN experiment \cite{Baker/2006} and $250 - 4000$ with respect to the best beam experiment \cite{Dress/1977}. Inserting the values for optimized conditions into Eq.\ (\ref{deltadn}) yields a nEDM sensitivity $\sigma(d_{\text{n}}) = 5 \times 10^{-28}$~e$\cdot$cm, i.e.\ a magnetic field sensitivity $4 B^*=3$~fT, assuming $100$~days of data taking and $\eta=0.75$. This matches the precision envisaged 
by future UCN experiments \cite{Grinten/2009,Masuda/2012a,Ito/2007,Altarev/2012,Baker/2011}. \\

%
%
%
% \section{Systematic Effects}
%
Finally, possible systematic effects are considered which could disguise a real or produce a false nEDM signal.
As already mentioned, the second order $v \times E$ term in Eq.\ (\ref{phases}) has to be smaller than the statistical sensitivity per day, i.e.\ $30$~fT, since the high voltage polarity will be reversed only a few times in 24 hours. 
This is achieved by a relative precision of the inverted fields $| \delta E_i /E_i| <5\times 10^{-4}$ and alignment accuracies e.g.\ $|\alpha_i |< 3$� and $| \delta\alpha_i |  < 0.5$�, assuming $v=500$~m/s, $E_i = 10$~MV/m and $B_0 = 200$~$\mu$T. The electric field precision can be assured by monitoring the high voltage and providing a corresponding mechanical stability. The angular alignment can be tested by applying an additional magnetic field in $x$ direction and minimizing the phase shift due to the first order $v \times E$-effect. 
The main systematic effect is due to changing magnetic field gradients.
Random fluctuations need to be monitored by means of gradiometers with a precision better than the statistical sensitivity of the neutron measurement. Gradients which are correlated with the orientation of the electric field, e.g.\ a magnetization of the mu-metal shield generated while reversing the high voltage polarity, have to be smaller than $3$~fT to achieve the final sensitivity goal. However, a false nEDM signal originating from $\delta B_g$ is inversely proportional to the electric field and can therefore be reduced up to a factor $10$ compared to nEDM searches with UCN.  
Another effect occurs, if there exist a net neutron velocity $v_x$ in $x$ and an electric field component in $y$ direction. This causes an effective $v \times E$-field along the $z$-axis. 
%The effect is suppressed below 1~fT, if $|v_x| < 1$~mm/s and the inclination angle between the electric and magnetic fields in $y$ direction is smaller than $0.1$� (with $E_i=10$~MV/m). 
The field is well suppressed, if e.g.\ $|v_x| < 2$~mm/s and the inclination angle between the electric and magnetic fields in $y$ direction is smaller than $0.2$� (with $E_i=10$~MV/m). 
An upper limit for the effect is obtained by intentionally increasing the inclination (add magnetic field in $y$ direction) or/and $v_x$ (shift beam apertures in $x$ direction).
Further, a geometric phase arises when a magnetic field component in $y$ direction exists with different values $B_{y,1}$ and $B_{y,2}$ at the entrance and exit of the electric field \cite{Commins/1991}. This yields in the adiabatic limit the phase
\begin{equation}
  \label{Geomphases}
    \Delta \Omega = \frac{\Delta B_y}{B_0^2} \cdot \frac{v}{c^2} \cdot (E_1 + E_2)
\end{equation}
which has to be added to Eq.\ (\ref{phases}), where $\Delta B_y= B_{y,1}-B_{y,2}$ and $\delta E_i \ll E_i$. The effect is, however, distinguishable from a nEDM signal by inverting the direction of the main magnetic field, if $\Delta B_y$ is caused by the coils producing $B_0$. 
An upper limit for $\Delta B_y$ which does not reverse with inverting the magnetic field can be determined by measuring at lower $B_0$ \cite{Abdullah/1999,Commins/1994}.
Moreover, a false signal caused by the geometric phase is suppressed below $5 \times 10^{-28}$~e$\cdot$cm, if $| \Delta B_y | < 20$~nT, assuming the aforementioned values for  $v$, $L$, $E_i$ and $B_0$.  \\

%
%
%
% \section{Conclusions}
%
In conclusion, a nEDM beam experiment has been reconsidered. A new concept exploits the advantage offered by a pulsed spallation source, to directly measure the so far limiting systematic $v \times E$-effect. The method is well superior to previous beam experiments and has the potential to significantly improve the present best measurement of the nEDM obtained with UCN. A statistical sensitivity of $5 \times 10^{-28}$~e$\cdot$cm can be achieved in $100$~days of data taking. \\

% by almost two orders of magnitude. \\
%
% \section{Acknowledgments}
%
The author gratefully acknowledges many useful discussions with Klaus Kirch, Peter B\"oni, Ben van den Brandt, Martin Fertl, Stefan Filipp, Patrick Hautle, Michael Jentschel, Jochen Krempel, Guillaume Pignol, Christian Schanzer, Torsten Soldner and Oliver Zimmer.

\appendix

\end{document}